\documentstyle[aps,amstex,amssymb,preprint,epsf]{revtex}
\begin{document}                                                       

\draft 

%\preprint{UASLP--IF--00--03}
                                            
%\title {Geomtric resonance and van der Waals forces}

%\author{B. I.\ Ivlev}

%\address{Department of Physics and Astronomy\\
%University of South Carolina, Columbia, SC 29208\\
%and\\
%\address{Instituto de F\'{\i}sica, Universidad Aut\'onoma de San Luis Potos\'{\i}\\
%San Luis Potos\'{\i}, S. L. P. 78000 Mexico}

%\date{\today}
%\maketitle

%\begin{abstract}

%\end{abstract} \vskip 1.0cm

%\pacs{PACS numbers: 42.50.Lc, 12.20.Ds, 03.70.+k} 

\narrowtext
{\bf Comment on ``Like-Charge Attraction and Hydrodynamic Interaction''}

The recent Letter [1] proposed a method to explain the observed micron-ranged attraction of colloidal particles on base 
of the macroscopic off-diagonal force in a hydrodynamic medium in the vicinity of the fluid boundary. The total effective 
interaction energy of two particles, separated by the distance $r$, in this approach is written as 
$U_{eff}(r)=U_{DLVO}(r)+U_{att}(r,h)$, where $U_{DLVO}$ is the repulsion DLVO potential and $U_{att}$ is the new attraction, proposed
in the Letter and produced by the motion of particles towards each other parallel to the fluid boundary, which is apart of 
the particle pair by a distance $h$. The repulsion and the attraction act together resulting in the minimum in the total 
potential $U_{eff}$, which, in the limit of non-small $\kappa r$, has the depth 
\begin{equation}
\label{2}
U_{min}\simeq u_{0}\hspace{0.1cm}\frac{\sigma_{g}}{\sigma_{p}}\exp[-(h-a)\kappa]\hspace{0.1cm}
f\left(\frac{r_{0}}{h},\kappa a,\kappa h\right)
\end{equation}
$r_{0}$ is the optimum distance, $\sigma_{p}$ and $\sigma_{g}$ are surface charge densities of particles and the glass plane, $a$
is the particle radius, $\kappa^{-1}$ is the Debye length, and $f$ is some function of the order of unity. $u_{0}$ is a scale 
of the electrostatic energy
\begin{equation}
\label{3}
u_{0}=\frac{(Ze)^{2}}{2a\varepsilon(1+\kappa a)^{2}}\simeq 10^{4}k_{B}T
\end{equation}

What the experiments say? Colloidal particles, confined between two glass plates, exhibit $U_{min}\simeq 0.2k_{B}T$ or less
for different ionic strength [2]. This experimental arrangement, studied in the other lab, gives 
$U_{min}\simeq (0.3 - 0.4)k_{B}T$ for different particle diameters and distances between plates [3]. The same type of the 
experimental arrangement of particles between two glass plates, studied in the third lab, gives $U_{min}\simeq 1.3k_{B}T$ or 
less for different ionic strength, screening length, and particle diameter [4]. Two particles in a one-wall geometry 
exhibit $U_{min}\simeq 0.7k_{B}T$ or less [5]. Polystyrene colloidal particles of a diameter $0.5\mu{\rm m}$ on the water-air 
interface exhibit $U_{min}\simeq 0.5k_{B}T$ at a center-to-center separation of $0.9\mu{\rm m}$ [6]. As one can see, despite of 
different conditions (even particles on the water surface), there is a very stable common feature of the all various 
experiments. The attraction minimum is always of the order of $k_{B}T$ and does not exceed $1.5k_{B}T$. This provides a 
hypothesis of some common mechanism of the micron-ranged attraction, which is responsible for the universality of 
$U_{min}$. A possible mechanism is rather an interaction mediated by thermal fluctuations in the system, resulting in the
sufficiently universal attraction of the order of $k_{B}T$ regardless a big dispersion of parameters [7].

How the method, proposed in the Letter [1], does fit the experimentally observed universality of $U_{min}$? According to 
Eq.~(\ref{2}), the big value of $u_{0}$, Eq.~(\ref{3}), should be compensated by the small exponent to obtain the 
experimental value $U_{min}\sim k_{B}T$ of Ref. [5]. This is possible, if to adjust $\sigma_{g}/\sigma_{p}$ to the set of other 
known parameters. The $h$-dependence of minima in Fig. 3 of Ref. [1] is reasonably proportional to $\exp(-\kappa h)$. For 
$h=2.0\mu{\rm m}$ the minimum depth becomes $U_{min}\simeq 6.8k_{B}T$ and it grows up fast with further decrease of $h$. 
Therefore, the theory, proposed in the Letter [1], is not consistent with the above universality, since it is based on the
accidental compensation of two big interactions, the repulsive $U_{DLVO}$ and the attractive $U_{att}$. In other words, the 
value of the order of $k_{B}T$ for $U_{min}$ is occasional for the proposed mechanism. The energy scale $k_{B}T$ is not 
specific in the theory, it is simply met on the way to the big $U_{min}\sim 10^{4}k_{B}T$, when $h$ decreases. Any attempt to 
apply the theory to experiments [2-4], as the authors suppose, will require a stable accidental compensation of the big 
$U_{DLVO}$ and $U_{att}$ down to $(0.2 - 1.3)k_{B}T$ in various experimental sets, which is impossible. The substantial point, 
giving an advantage to the presented theory over the universality mechanism in explanation of the results [5], would be an
observation of big $U_{min}/k_{B}T$. But it was not observed, the biggest measured $U_{min}$ in Ref. [5] is of $0.7k_{B}T$. This
says that the attraction in Ref. [5] is rather of the same common mechanism like in Refs. [2-4, 6]. The similarity of the 
two calculated curves with experimental ones is not a firm argument, because one-parameter adjustment of the power law 
curve $U_{att}$ to the known exponential curve $U_{DLVO}$ can easily fit a region near the minimum of the curve with
$h=2.5\mu{\rm m}$. The second curve $(h=9.5\mu{\rm m})$ is simply DLVO.

According to the above arguments, the proposed mechanism is very unlikely relevant in explanation of the observed 
micron-ranged attraction of colloidal particles, since the essential aspect of this attraction, $U_{min}\sim k_{B}T$, is not 
accounted in the theory as a stable feature. The particular experimental result is explained on base of an accidental 
parametric fit. Nevertheless, the method of the off-diagonal hydrodynamic interaction is interesting and it may be 
applicable to experiments, which will show a big value of $U_{min}/k_{B}T$.\\

Boris Ivlev

Department of Physics and Astronomy,

University of South Carolina, Columbia SC 29208;

Universidad Autonoma de San Luis Potosi,

Instituto de Fisica, A.Obregon 64, San Luis Potosi,

78000 S.L.P. Mexico\\


[1] T. M. Squires and M. P. Brenner, Phys.Rev.Lett. {\bf 85}, 4976 (2000)

[2] G. M. Kepler and S. Fraden, Phys.Rev.lett. {\bf 73}, 356 (1996)

[3] M. D. Carbajal-Tinoco, F. Castro-Roman, and J. L. Arauz-Lara, Phys.Rev.E {\bf 53}, 

3745 (1996)

[4] J. C. Crocker and D. G. Grier, Phys.Rev.Lett. {\bf 77}, 1897 (1996)

[5] A. E. Larsen and D. G. Grier, Nature (London) {\bf 385}, 230 (1997)

[6] J. Ruiz-Garcia, private communication.

[7] B. I. Ivlev, preprint cond-mat/0004221 (2000)

%\section{ACKNOWLEDGMENT} 

%\begin{references}

%\bibitem{}

%\end{references}

%\newpage
%\begin{figure}[p]
%\begin{center}
%\vspace{2cm}
%\leavevmode
%\epsfxsize=\hsize
%\epsfxsize=12cm
%\epsfbox{figu1.eps}
%\vspace{2cm}
%\caption{AAAAA}
%\label{fig1}
%\end{center}
%\end{figure}

\end{document}